\documentclass[a4paper,12pt]{article}

\usepackage{graphicx}

\begin{document}

\begin{center}
{\large\bf Regge phenomenology of photoproduction of
$\pi^-\Delta^{++}$ and scaling with saturation of trajectory}

{Byung-Geel Yu$^{1,\P}$}, {Kook-Jin Kong$^{1}$}

$^1${Korea Aerospace University, address, 10540, Goyang, Korea}

%$^2${korea Aerospace University, address, 10540, Goyang, Korea}

$^\P${E-mail: bgyu@kau.ac.kr}
\end{center}

\centerline{\bf Abstract} In this work we investigate the reaction
$\gamma p\to\pi^-\Delta^{++}$ in the Reggeized model for
$\pi(138)+\rho(775)+a_2(1320)$ exchanges  in the $t$-channel. For
a convergence of the reaction cross section at high energies the
minimal forms of proton and $\Delta^{++}$ exchanges are introduced
in the direct and crossed channels for gauge invariance of $\pi$
Regge-pole exchange. The role of spin-2 tensor meson $a_2$ is
found to be crucial to agree with existing data at high energies.
Electromagnetic multipoles of $\Delta^{++}$ baryon are analyzed in
the $\Delta$ resonance region. Based on the constituents counting
rule the scaled differential cross section at $E_\gamma=4$ GeV is
reproduced with the Regge trajectory saturated at large momentum
transfer $-t$.
\\


\section{Introduction}

It is known that the Regge model describes hadron reactions over
the resonance region without either cutoff functions or
fit-parameters \cite{guidal}. Meanwhile, photoproduction of pion
with $\Delta$ baryon in the final state is interesting because it
provides information about the internal excitation of nucleon by
electromagnetic (EM) probe. In this work we investigate the
reaction $\gamma p\to\pi^-\Delta^{++}$ to understand the
production mechanism of the reaction at high energies and the
scaling of differential cross section at wide angles by analyzing
experimental data. There are plenty of data on the reaction cross
sections measured up to 16 GeV photon energy
\cite{wu,ballam,anderson}. To date these are analyzed based on the
$\pi+\rho$ exchanges. However, theoretical analysis of the
reaction at such high energy has not been completed yet, because
the $\pi^-\Delta^{++}$ photoproduction is known to diverge due to
the divergence of proton and $\Delta^{++}$ poles at high energies.
For this issue we followed a special gauge prescription in
previous study \cite{bgyu2} for the convergence of proton and
$\Delta$ propagations for $\pi$ exchange. Nevertheless, the data
at high energy such as $E_\gamma\approx 16$ GeV cannot be
reproduced without further contribution of $t$-channel meson
exchanges. To improve such a lack in theoretical predictions
existing  we consider the tensor meson $a_2$ exchange and examine
its role in the reaction process. At low energy, the static
properties of $\Delta$ baryon are important to study the internal
structure of hadrons based on the quark dynamics. We calculate
contributions of electromagnetic (EM) multipole moments of
$\Delta^{++}$ baryon to the total cross section with the EM form
factors of $\Delta^{++}$ fully considered. On the other hand, pQCD
predicts quark evidences in the photoproduction of hadron,
manifesting themselves through the scaling of differential cross
section at wide angles. Within the present framework we
investigate scaling of the differential cross section for
$\pi^-\Delta^{++}$ process by considering the saturation of the
Regge trajectory at large momentum transfer $-t$.

\section{The model and observables}

The reaction $\gamma p\to\pi^-\Delta^{++}$ at high energies is
dominated by one pion exchange through the Drell process, because
the reaction cross section measured in experiments shows a steep
decease as the reaction energy increases. To render the $\pi$
exchange in the process gauge invariant the charge couplings of
the $s$-channel proton and $u$-channel $\Delta^{++}$ poles are
introduced by following the charge conservation,
\begin{eqnarray}\label{pi-regge}
&&i{M}_{\pi N\Delta}=\frac{f_{\pi
N\Delta}}{m_{\pi}}\bar{u}_{\nu}(p')\biggl[
q^\nu\frac{2p\cdot\epsilon+\rlap{/}{k}
\rlap{/}{\epsilon}}{s-M^{2}_{N}}e_N
%\nonumber\\&&
+e_\Delta\frac{q_{\mu}} {u-M^{2}_{\Delta}}\left(2p'\cdot\epsilon
g^{\mu\nu}+\sum_{i}G_i^{\nu\mu}(p',k,\epsilon)\right)\nonumber\\&&
+e_{\pi}\frac{2q\cdot\epsilon }{t-m^{2}_{\pi}}(q-k)^\nu
-e_\pi\epsilon^\nu\biggr]u(p),
\end{eqnarray}
which is then used to reggeize the $\pi$ exchange as
\begin{eqnarray}
{\cal M}_\pi=M_{\pi N\Delta}\times(t-m_\pi^2)\times
{\pi\alpha_\pi'(1+e^{-i\pi\alpha_\pi(t)})\over2\Gamma
(\alpha_\pi(t)+1)\sin\pi\alpha_\pi(t)}\left({s\over
s_0}\right)^{\alpha_\pi(t)}
\end{eqnarray}
with the trajectory $\alpha_\pi(t)=0.7(t-m_\pi^2)$ and the
collection of transverse terms, i.e., $k_\mu
G_i^{\nu\mu}(p',k,k)=0$ in the $\Delta$ pole, which contains the
divergent terms at high energy. The minimal gauge requires removal
of the transverse components $\rlap{/}k\rlap{/}\epsilon$ and
$G_i^{\nu\mu}(p',k,k)$ in Eq. (\ref{pi-regge}) by redundancy for
gauge invariance of $\pi$ Regge-pole \cite{bgyu2}.
\bigskip

%%%%%%%%%%%%%%%%%%%%%%%% FIG 1 and FIG 2
%\vspace*{-1.cm}
\begin{figure}[htb]
%\begin{center}
\begin{minipage}[t]{75mm}
\includegraphics[width=16.4pc]{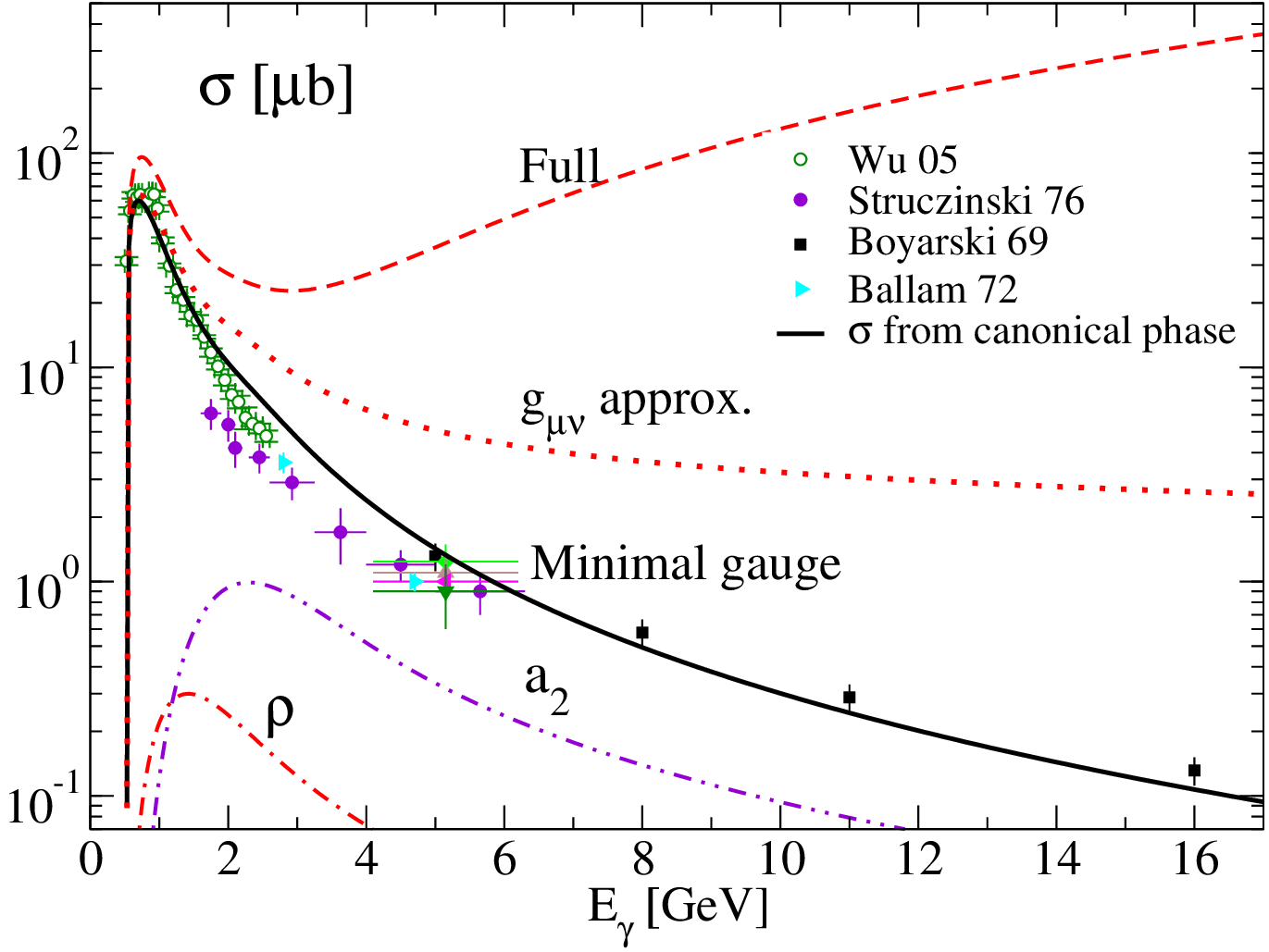}
%\framebox[79mm]{\rule[-26mm]{0mm}{52mm}}
\caption{Total cross section from the full and minimal gauge up to
$E_\gamma=16$ GeV \cite{bgyu2}. } \label{fig1}
\end{minipage}
\hspace{\fill}
\begin{minipage}[t]{75mm}
\includegraphics[width=16.4pc]{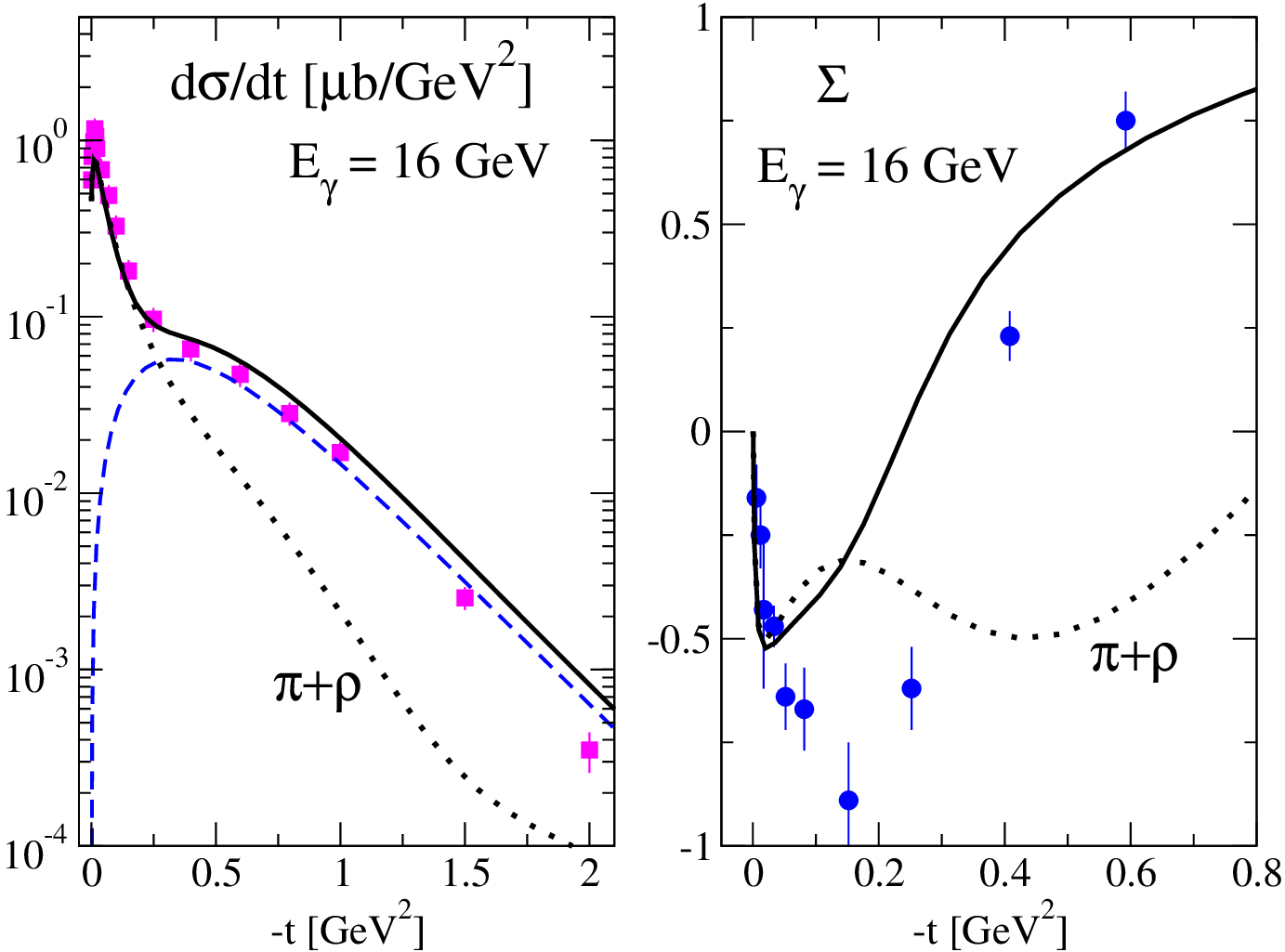}
%\framebox[74mm]{\rule[-26mm]{0mm}{52mm}}
\caption{ $d\sigma/dt$ and $\Sigma$ at $E_\gamma=16$ GeV.}
\label{fig2}
\end{minipage}
%\end{center}
\end{figure}
%%%%%%%%%%%%%%%%%

We now ask what is the next contribution to pion exchange in the
$t$-channel.  For this question we investigate roles of  vector
meson $\rho$ and tensor meson $a_2$ Regge pole exchanges
independently, although these are exchange degenerate in the
$\rho$ trajectory. For the $\rho$ exchange we consider the well
known form of the $\rho N\Delta$ coupling here. But for the case
of $a_2N\Delta$ there is no information about either the coupling
constants or the interaction Lagrangian at present. In this work
by considering spin-parity conservation of $a_2N\Delta$ coupling
we construct a new Lagrangian of the form as
\begin{eqnarray}\label{a2}
&&{\cal L}_{a_2N\Delta}=i{f_{a_2N\Delta}\over
m_{a_2}}\bar\Delta^\lambda\left(g_{\lambda\mu}{\partial_\nu}
+g_{\lambda\nu}{\partial_\mu}\right) \gamma_5 N
{a_2}^{\mu\nu}+\mathrm{h.c.}\,.
\end{eqnarray}
Here we use the identity to determine the $f_{a_2N\Delta}$
coupling constant \cite{goldstein},
\begin{eqnarray}\label{id}
{f_{a_2 N\Delta}\over m_{a_2}}=-3{f_{\rho N\Delta}\over
m_{\rho}}\,,
\end{eqnarray}
based on the fact that the coupling constant
$g_{\gamma\pi\rho}=0.224$ is the same order of magnitude as
$g_{\gamma\pi a_2}=-0.276$ \cite{bgyu2}.

Fig. \ref{fig1} shows the results in total and differential cross
sections and beam polarization up to $E_\gamma=$16 GeV. We use the
radiative decay coupling constants from the Particle data Group
and the meson-baryon coupling constants $f_{\pi N\Delta}=2.0$, and
$f_{\rho N\Delta}=8.57$ with the trajectories and phases as given
in Ref. \cite{bgyu2}. Recall that the $a_2 N\Delta$ coupling
constant is not a free parameter, but  determined once the $\rho
N\Delta$ coupling is chosen. The total cross section with the
minimal gauge shows a good behavior of convergence at high
energies. The role of $a_2$ exchange is clear rather than the
$\rho$, and even indispensable to agree with the differential
cross section and beam polarization at 16 GeV in Fig. \ref{fig2},
as compared to the dotted curves which are the cases without $a_2$
exchange.

\begin{figure}[htb]
%\begin{center}
\begin{minipage}[t]{75mm}
\includegraphics[width=16.6pc]{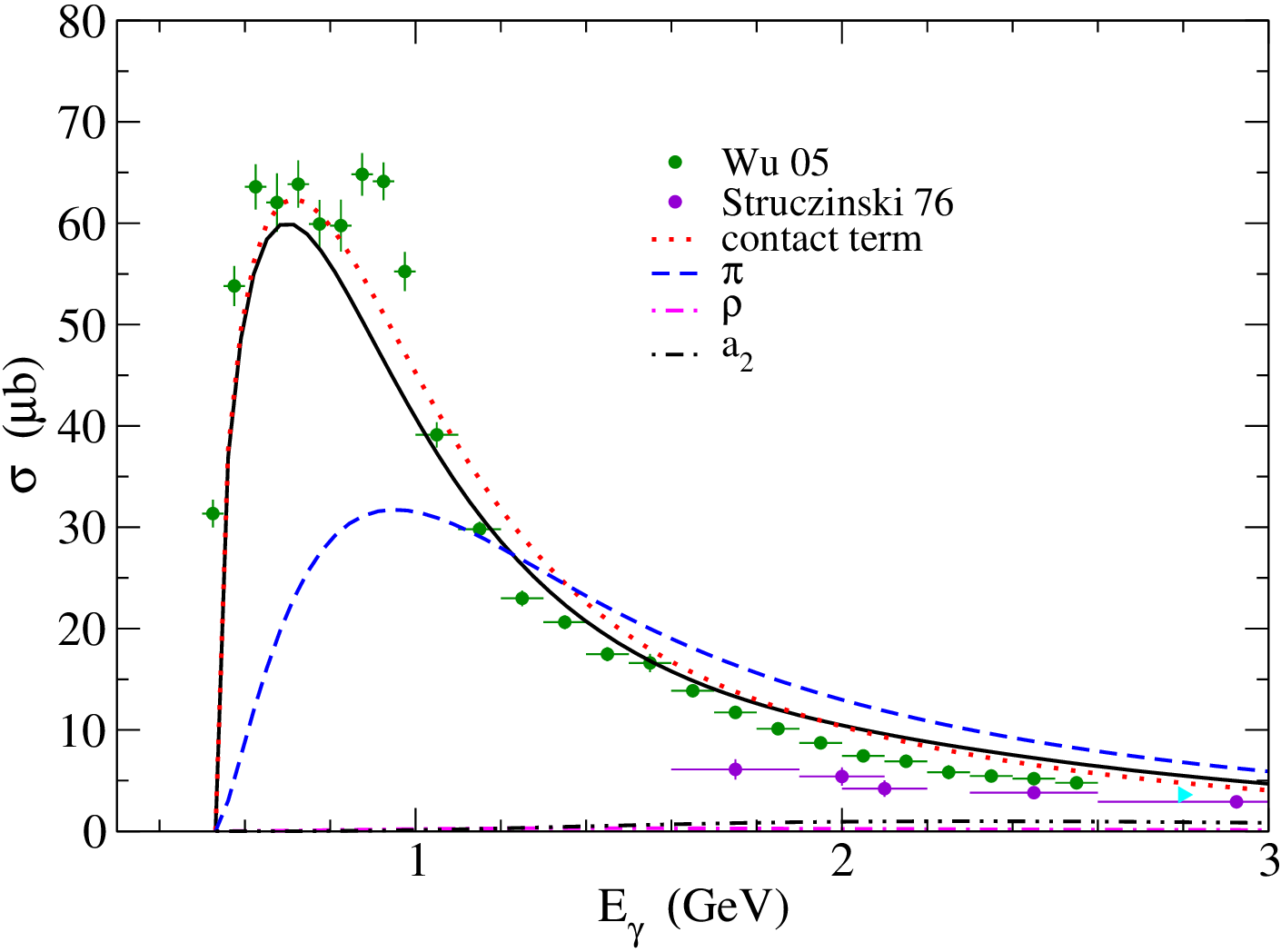}
%\framebox[79mm]{\rule[-26mm]{0mm}{52mm}}
\caption{Total cross section  at low energy from the minimal
gauge. } \label{fig3}
\end{minipage}
\hspace{\fill}
\begin{minipage}[t]{70mm}
\includegraphics[width=16pc]{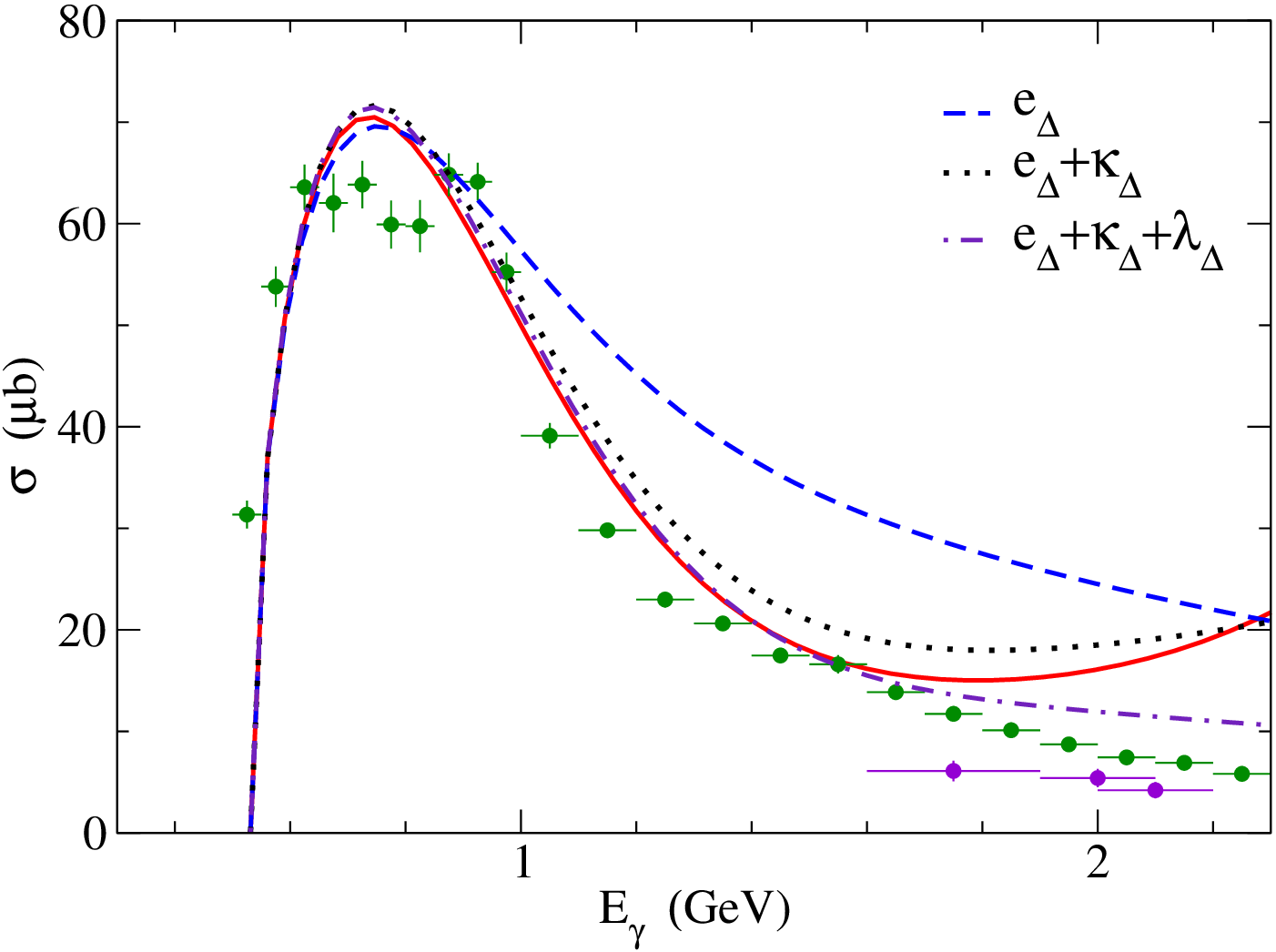}
%\framebox[74mm]{\rule[-26mm]{0mm}{52mm}}
\caption{EM multipole moments of $\Delta^{++}$ with
$\Gamma_{\gamma\Delta\Delta}$ vertex in Eq. (\ref{gamma4}).}
\label{fig4}
\end{minipage}
%\end{center}
\end{figure}

Fig. \ref{fig3} shows the validity of the present model at low
energy with only the charge couplings as discussed above. But the
model could also be applied to study electromagnetic (EM)
multipole moments of $\Delta^{++}$ as we consider the
${\gamma\Delta\Delta}$ coupling vertex to have the four EM moments
\begin{eqnarray}
    &&\epsilon_\mu\Gamma_{\gamma\Delta\Delta}^{\lambda\mu\sigma}(p',k,p)=
-e_\Delta(g^{\lambda\sigma}\rlap{/}\epsilon-\epsilon^\lambda\gamma^\sigma
    -\gamma^\lambda \epsilon^\sigma +\gamma^\lambda\rlap{/}\epsilon
    \gamma^\sigma)%\nonumber\\&&\hspace{2cm}
+{e\over4M_\Delta}\left( \kappa_\Delta\,g^{\lambda\sigma}
    +\chi_\Delta{k^\lambda
        k^\sigma\over4M^2_\Delta}\right)\left[\rlap{/}{\epsilon},\,\rlap{/}{k}\right]
    \nonumber\\&&\hspace{3.5cm}
-{e\lambda_\Delta\over
        4M^2_\Delta}\left[k^\lambda k^\sigma
    \rlap{/}\epsilon-{1\over2}\rlap{/}k(\epsilon^\lambda
    k^\sigma+\epsilon^\sigma k^\lambda) \right],\label{gamma4}
\end{eqnarray}
with $e_{\Delta^{++}}=2$, $\kappa_{\Delta^{++}}=4.34$,
$\chi_{\Delta^{++}}=12.34$, and $\lambda_{\Delta^{++}}=6.8$ taken
from Ref. \cite{azizi}. For a better agreement with data we use
$f_{\pi N\Delta}= 1.7$ and $f_{\rho N\Delta}=5.5$ in this case and
present the result  in Fig. \ref{fig4} where the propagations of
proton with $\kappa_p=1.79$ and $\Delta^{++}$ with spin-3/2
projection operator are fully considered. Of course, the cross
section without the minimal gauge should diverge, as can be seen
over $E_\gamma\approx 1.6$ GeV. Nevertheless, we notice that  our
model works good enough to test these EM multipoles below the
region convergent.
\bigskip

%%%%%%%%%%%%%%%%%%%%%%%% FIG 3 and FIG 4
%\vspace*{-1.cm}
%%%%%%%%%%%%%%%%%

Let me now discuss the reaction at large momentum transfer.
According to the dimensional analysis based on the pQCD
calculation, differential cross sections at the production angle
$\theta=90^\circ$ become angle independent and the energy
dependence is given by the powers of energy squared, $s^{2-n}$.
Here $n$ is total number of particles participating in the
reaction. In photoproductions of hadrons, $n=9$ and the
differential cross section multiplied by the factor of $s^7$,
therefore, leads to an energy independence, i.e., scaling.

In Regge models the energy dependence of a differential cross
section for the exchange of $\alpha(t)$ trajectory is given by
\begin{eqnarray}\label{dcs}
{d\sigma\over dt}\simeq s^{2\alpha(t)-2}.
\end{eqnarray}
At wide angles, or alternatively large $-t$, the trajectory of
Regge-pole should  saturate to a limiting value and it is known
that  $\alpha(t)\to -1$ for the case of meson \cite{collins}. In
such kinematical conditions the differential cross section in Eq.
(\ref{dcs}) is expected to behave as $s^{-4}$ in energy and, as a
result, it should diverge with $s^3$ when scaled by $s^7$ factor.
For this reason we need to introduce a form factor  to suppress
it.

For the saturation of a linear trajectory $\alpha(t)$ we use a
simple parameterization of the square root function \cite{collins}
\begin{eqnarray}\label{traj}
\alpha^*(t)=c_1+c_2\sqrt{t_1-t},
\end{eqnarray}
where the coefficients $c_1$ and $c_2$ are determined by the
boundary conditions $\alpha^*(t_0)=\alpha(t_0)$ and
$d\alpha^*(t_0)/dt=d\alpha(t_0)/dt$ at the saturation  point $t_0$
we choose to make the trajectory saturating to $-1$ with
$t_0<t_1$. Then $t_1$ is the initial point of the square root
function we take for calculation.

\begin{figure}[htb]
%\begin{center}
\begin{minipage}[t]{75mm}
\includegraphics[width=16.6pc]{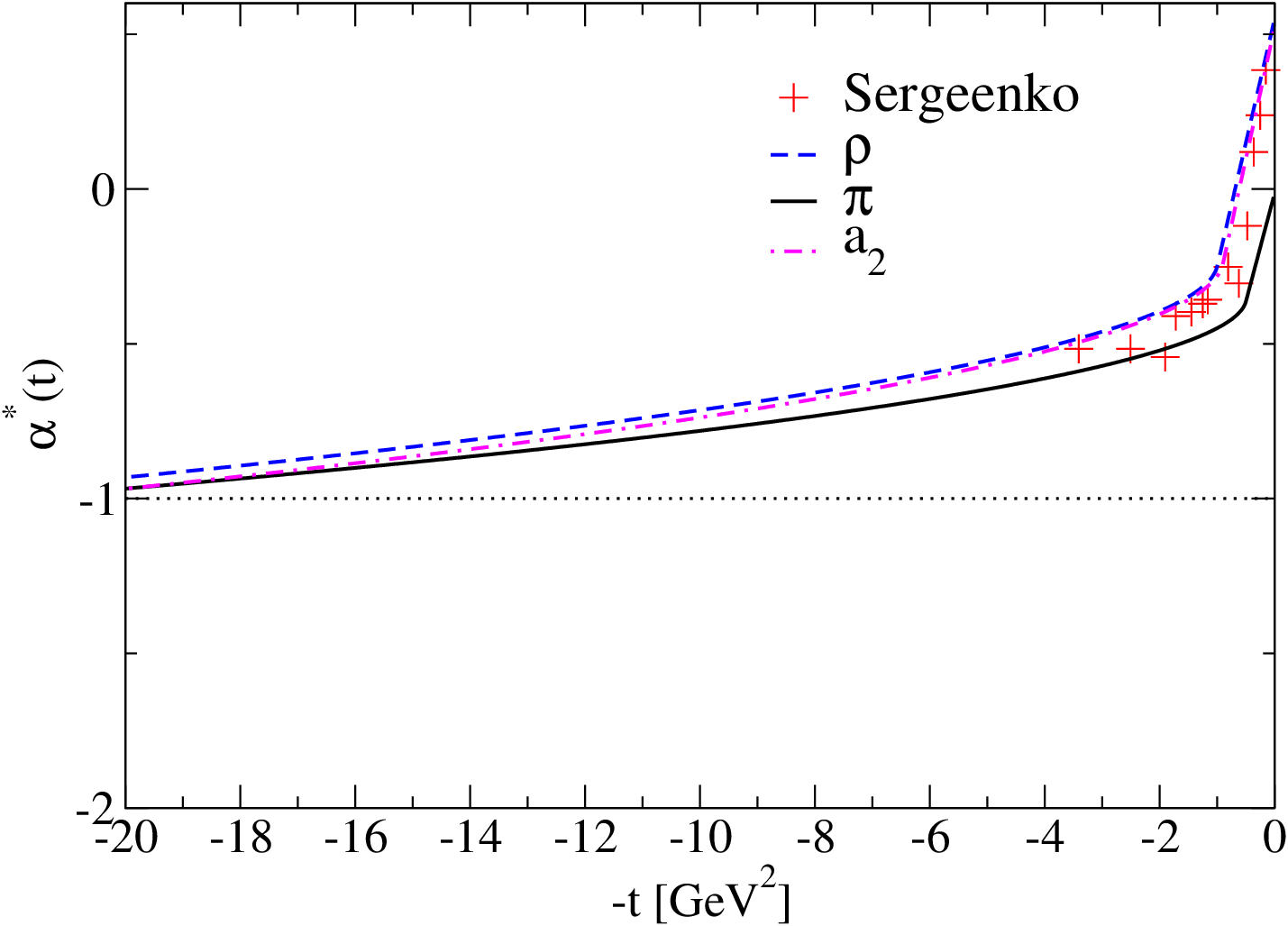}
%\framebox[79mm]{\rule[-26mm]{0mm}{52mm}}
\caption{Saturation of trajectory following Eq. (\ref{traj}). }
\label{fig5}
\end{minipage}
\hspace{\fill}
\begin{minipage}[t]{75mm}
\includegraphics[width=16.6pc]{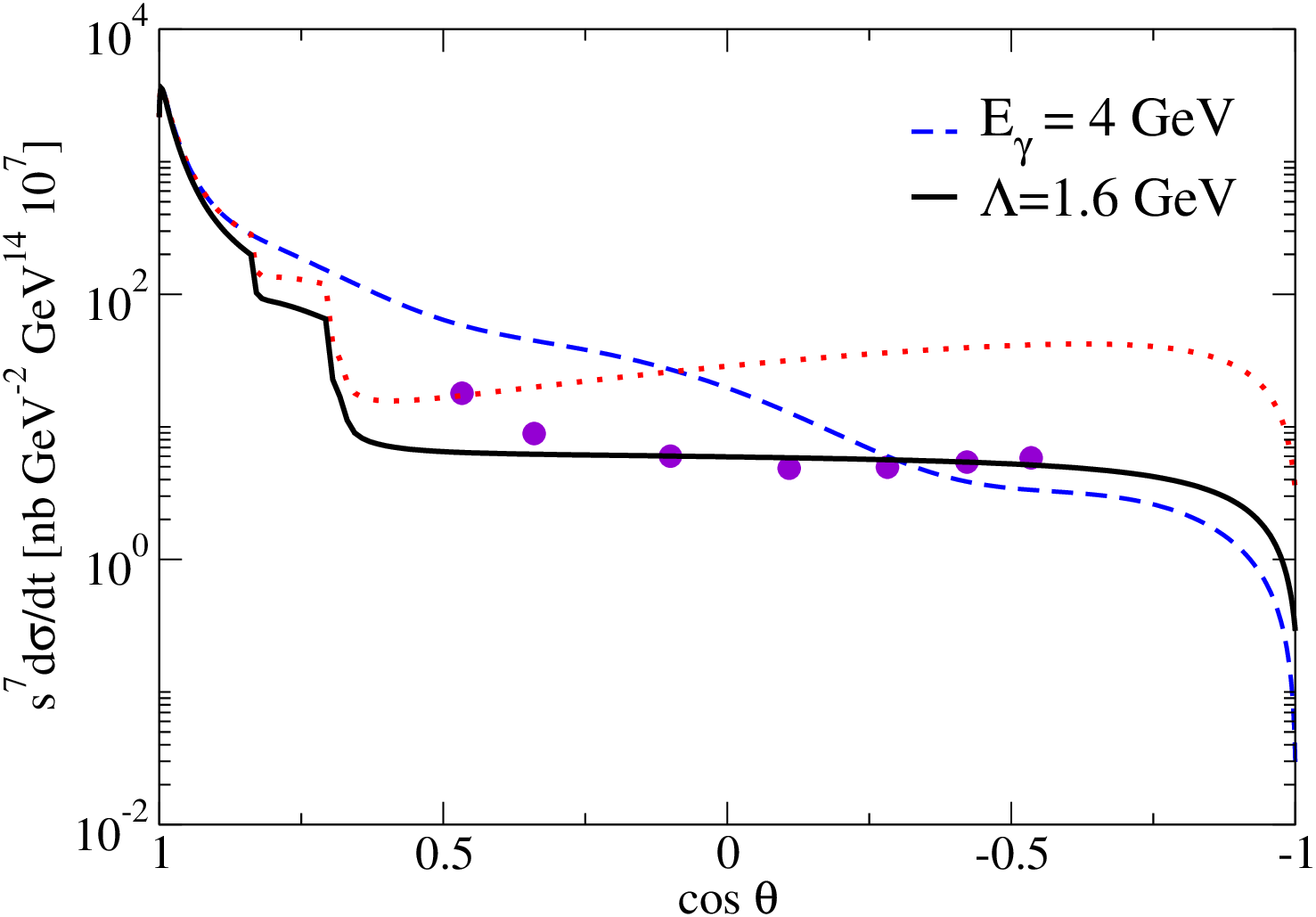}
%\framebox[74mm]{\rule[-26mm]{0mm}{52mm}}
\caption{Scaled differential cross section. Data are taken from
Ref. \cite{anderson}.} \label{fig6}
\end{minipage}
%\end{center}
\end{figure}

In Fig. \ref{fig5} we show the saturation of $\pi$ trajectory by
choosing $t_0=-0.5,\,t_1=-0.49$, of $\rho$ by
$t_0=-1,\,t_1=-0.99$, and $a_2$ by $t_0=-0.9,\,t_1=-0.89$ in unit
of GeV$^2$, which lead them definitely saturating to $-1$ up to
$-t = 20$ GeV$^2$. The (red) crosses calculated by  Sergeenko for
the $\rho$ trajectory are given for comparison \cite{sergeenko}.
For $\pi^-\Delta^{++}$ photoproduction the result of the scaled
differential cross section is presented in Fig. \ref{fig6} to
compare with data at $E_\gamma=4$ GeV from the SLAC experiment. We
used the form factor of the monopole type
\begin{eqnarray}\label{monopole}
F(t)=\left(1-{t\over\Lambda^2}\right)^{-1},
\end{eqnarray}
as an overall factor of the production amplitude in Eq.
(\ref{dcs}), and obtained a good fit to experimental data at the
cutoff mass 1.6 GeV.  The dashed curve is from the trajectory
unsaturated and the dotted one is the case of the trajectory
saturated, but without the form factor. Thus, it shows a
divergence over data, as expected.
\bigskip
\bigskip

%%%%%%%%%%%%%%%%%%%%%%%% FIG 5 and FIG 6
%\vspace*{-1.cm}
%%%%%%%%%%%%%%%%%

\section{conclusion}

In summary, we investigated the reaction $\gamma
p\to\pi^-\Delta^{++}$ from threshold to the Regge realm based on
the Reggeization of the $t$-channel exchanges.
At high energies up to $E_\gamma=16$ GeV a good convergence of
cross section is obtained with the minimal gauge for $\pi$
exchange, and the role of tensor meson $a_2$ is found to be
indispensable to reproduce the data on differential cross section
and beam polarization. The present approach yields an agreement
with data on total cross section in the resonance region, in which
case the role of $\Delta^{++}$ EM multipole moments is further
analyzed with the $\gamma\Delta\Delta$ vertex fully considered. A
saturation of the trajectory is applied to describe the scaled
differential cross section at large $-t$. Such numerical
consequences in good agreement with existing data confirm the
validity of the present framework with  great potential in the
study of hadron reactions up to high energy and wide angle.

\section*{Acknowledgment}
This work was supported by the grant NRF-2017R1A2B4010117 from the
National Research Foundation of Korea.

\end{document}